# Ion assisted structural collapse of a single stranded DNA: a molecular dynamics approach

Soumadwip Ghosh, Himanshu Dixit, Rajarshi Chakrabarti[*]

*Department of Chemistry, Indian Institute of Technology, Powai, Mumbai – 40076, India.*

**Abstract:** The structure and dynamics of negatively charged nucleic acids strongly correlate with the concentration and charge of the oppositely charged counterions. It is well known that the structural collapse of DNA is favoured in the presence of additional salt, a source of excess oppositely charged ions. Under such conditions single stranded DNA adopts a collapsed coil like conformation, typically characterized by stacking base pairs. Using atomistic molecular dynamics simulation, we demonstrate that in the presence of additional divalent salt ($MgCl_2$) single stranded DNA with base sequence 5'-CGCGAATTCGCG–3'(Dickerson Drew dodecamer) initially collapses and then expands with increasing salt concentration. This is due to the overcharging induced DNA chain swelling, a dominant factor at a higher divalent salt concentration. In a nutshell, our simulations show how in the presence of divalent salt, non-sequential base stacking and overcharging competes and affect single stranded DNA dynamics unlike a monovalent salt.

## I. INTRODUCTION

Nucleic acids have crucial biological functions in gene storage, replication, repair and gene regulation[1,2]. Single stranded DNA (ss-DNA) molecules are important intermediates in most of the DNA mediated metabolisms. ss-DNA in their native state is inherently polyanionic in nature because of its negatively charged phosphate backbones. In single stranded DNA this intra strand repulsion between individual phosphate residues renders the overall system stable against folding. On the other hand, charged cations (e.g $Na^+$, $Mg^{2+}$) screen this electrostatic repulsion and help in the structural collapse of single stranded DNA. ss-DNA-ion and ss-DNA-water interactions attribute to the conformational flexibility of the chain and have been addressed in experimental as well as simulation studies. In a recent experimental study, the dependence of persistence length of DNA and RNA on salt concentration has been studied by Kuznetsov et al.[3] by using equilibrium DNA hairpin melting profiles. Bizarro et al.[4] determined the same parameters in the presence of $Na^+$ using atomic force microscopy[5]. Holm[6] showed the formation of a strongly correlated liquid driven by the weak Columbic interaction of ions along ss-DNA chain by using radial distribution calculations. Amongst different models prescribed for ion-binding fluctuations, the tight bound ion (TBI) is most successful in predicting the roles of monovalent/ divalent cations in stabilizing RNA hairpins[7,8] and RNA tertiary structures[9]. The site-specific hydration behaviour of single stranded DNA has been probed by absorption spectroscopy[10,11]. Conformational flexibility of oligo-dT is found in the range between 10 and 70 nucleotides from 0.025 to 2M NaCl using FRET[12]. Mills et al. examined the flexibility of oligonucleotides in the presence of $Mg^{2+}$ by means of transient electric birefringence[13]. Sen and Nilsson[14] have demonstrated the configurational preference of DNA and RNA single strands as a function of bridging water-backbone interaction and base pair stacking. Furthermore, the elastic behaviour of an ss-DNA has also been studied both by atomistic simulation as well as single-molecule experiments[15].

The sequence specific distributions of purine and pyrimidine nucleobases in the native states of DNA govern the conformational properties of the ss-DNA. However, the ss-DNA adopts a collapsed coil like conformation in the presence of ions and the distribution of stacked base pairs gets altered significantly[16]. In the collapsed coil like state, bases which do not obey the original sequence connectivity, may get stacked with each other because of their close proximity unlike the native state of the DNA. This non-sequential mode of base pair stacking[16] is one of the main stabilizing forces in the collapsed form of an ss-DNA as the interaction energy (both Columbic as well as van der Waals) between such a stacked base pair contributes appreciably to the overall attractive potential of the system. The conformational fluctuations of the DNA can also be characterized by considering the dynamics of the ions and solvent molecules around since the flexibility of an ss chain is sensitive towards its 'ionic environment'. Therefore, the study of ionic charge, size and concentration are of huge importance in determining the flexibility of the overall DNA. Both theoretical calculations as well as computer simulations predict that multivalent ions are more efficient in inducing the structural collapse of polyelectrolytes[17]. In most of these computational methods, a coarse-grained[18] model of an ss-DNA was systematically analyzed by Monte Carlo simulation[19]. Although the use of coarse-graining reduces the number of atoms and simplifies the system, it misses out chemical details which can only be captured in all atom simulation. As for example, the structural flexibility of the ss-DNA as a function of increased positive charge on the ion and its concentration was well addressed by Wang & co-workers[20] but further information about the base pair stacking both in the native as well as collapsed state of the DNA could not be extracted due to the coarse-grained formalism. An alternate atomistic molecular dynamics simulation would be computationally expensive but worth employing as it would be extremely useful in analyzing the conformational changes of the molecule and the preferential affinity of DNA bases and backbone for ions and the solvent. To the best of our knowledge, there have not been many attempts to investigate the dynamic of ss-DNA in the presence of moderate to highly concentrated divalent cation ($Mg^{2+}$) using atomistic simulations[21]. In this paper, using atomistic simulations we investigate the DNA-$Mg^{2+}$ and DNA-water interactions and its implications on the conformational changes associated with DNA collapse. We primarily address two issues. Firstly, we show that the dynamic structural changes of the DNA chain do not necessarily exhibit a monotonic dependence on the $Mg^{2+}$ concentration which indicates DNA chain expansion owing to overcharging[22]. Secondly, we show how non-sequential base stacking actually contributes to the DNA collapse. For our study we take a single stranded Dickerson Drew dodecamer[23] with base sequence 5'- CGCGAATTCGCG –3' and all the simulations are carried out at 300K. A trajectory of the DNA structure is obtained after a 50 ns simulation and the calculations of different parameters reveal the discrete patterns of structural collapse of ss-DNA when different counterion concentrations are involved under otherwise similar MD simulation conditions. MD simulations of the same ss-DNA at different $Mg^{2+}$ concentrations (0.01, 0.05, 0.1, 0.3 and 1M) reveal a trend in structural changes, qualitatively different to that observed in the case of a monovalent species such as $Na^{+}$[16]. A combined analysis of the data reported in this paper, may be helpful in comparing the patterns encountered in the structural collapse of the DNA at different concentrations of a divalent counterion such as $Mg^{2+}$. The paper is arranged as follows: In section II we present the simulation details and section III deals with the simulation results. The paper ends with the conclusions in section IV.

## II. DNA model and simulation details

All the molecular dynamics simulations are performed using GROMACS 4.5.6[24] with the all atom CHARMM[25] force field. An explicit SPCE water model[26] is used to solvate the DNA, the PDB file containing the initial coordinates of the DD dodecamer is downloaded (PDB ID 436D)[27] from Brookhaven protein data bank. Then the base sequence of the DNA oligomer is obtained by separating the complementary B chain from the A chain and capping the 5' and 3' phosphate end groups. Hydrogen atoms are then inserted in the initial DNA configuration using the freely available package 3DNA[28]. Subsequently, the entire DNA molecule is kept inside a cubic box of length 10 nm having 31263 explicit SPCE water molecules. The net charge of the native system is found to be -11 and hence 6 $Mg^{2+}$ and 1 $Cl^-$ ions are added for charge neutralization. The charge neutralized system corresponds roughly to 0.01M $MgCl_2$ concentration. The addition of co-ions ($Cl^-$) while adding a large number of DNA counterions controls high positive charge build up during simulations[29] and the systems are neutralized accordingly while setting up the system at higher concentrations of the additional salt. The time evolution of the distribution of the randomly placed $Mg^{2+}$ for each concentration has been depicted in the **supplementary section, FIG.S1**. However, it may be noted that the distributions of ions at each salt concentration are influenced strongly by the statistical sampling time (50 ns in the present study) and the manner in which they are introduced into the system at the beginning of each simulation[30].

Energy minimization of the system is a key simulation step since it takes care of any kind of unfavourable strain on the system. We employ the steepest descent algorithm[15] in order to converge the potential energy of the system. Next, the system is equilibrated in constant volume and temperature ensemble (NVT) performed at a temperature of 300K for 1 ns. Once the system reaches the desired temperature, it is then equilibrated using the isobaric-isothermal ensemble (NPT) for 2 ns at this temperature. This step is carried out at a constant pressure of 1 bar using Parrinello-Rahman borostat[31] and the v-rescale thermostat[32] is used to keep the temperature of the system constant at 300 K and the system configuration is updated by GROMACS using the leap frog integrator[33]. After the completion of the equilibration steps, the production MD run is started for 50 ns. We would like to mention that accurate statistical sampling of systems like ours which stay away from equilibrium is very difficult. One can reproduce a statistical ensemble correctly if the sampling time is sufficiently long. However the main purpose of this work is to extract qualitative trends of structural DNA collapse in the presence of a divalent salt and the factors influencing the dynamics. Since, most of the initial tendencies of the dynamics parameters can already be obtained from a short simulation of 50 ns we did not carry out longer simulations in order to improve the statistical sampling of the system. The average temperature of the equilibrated trajectory is found to be 300.012K. The entire production MD is carried out with a time step of 2 fs and the information regarding trajectory, velocity and energy are stored after each 2 ps for analysis. The minimum image convention[34] is used to calculate the short ranged Lennard–Jones interactions. The spherical cut-off distance for both electrostatic as well as van der Waals forces is kept 1 nm. The SHAKE[35] algorithm is used to impose a holonomic constrain on the equilibrium bond distance of the SPCE water molecules. The long range electrostatic interactions are calculated using the particle mesh Ewald[36] method. The specific base pairs of interest are tagged as energy groups while starting each MD run.

**III. Simulation results**

**A. Structural features of DNA in presence of 0.01M MgCl$_2$**

Several configurations of the ss-DNA molecule as obtained from the simulated trajectory at different time intervals are shown in Fig. 1. Since we are interested in probing conformational flexibility of the DNA chain as a function of counterion concentration, it is of primary importance that we address the collapsed form of the DNA. VMD[37] screenshots of the DNA at different simulation time intervals are shown here which point out the fact that the molecule undergoes significant conformational changes as the time progresses during MD run.

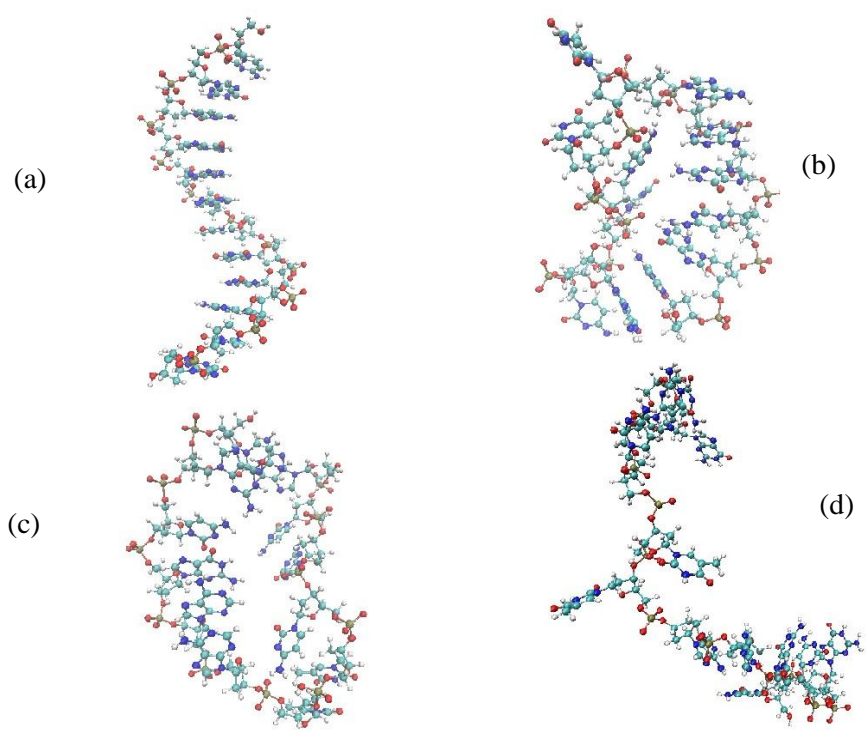

**FIG 1:** Screenshots of the (a) native and (b) – (d) representative configurations of the ss-DNA in the presence of 0.01M MgCl$_2$ at 20, 30 and 50 ns of the simulation respectively.

In practice, root mean square deviation (RMSD) is used to estimate the conformational changes of the DNA chain from its native structure. **FIG.2.a** depicts the variation of RMSD values with simulation time for DNA bases and the backbones. The estimation of RMSD covered the whole simulated trajectory excluding hydrogen atoms. The time evolution and the distribution of RMSD values are the signatures of the collapsed state of the DNA (**FIG.2.a& 2.b**). It can be seen from the plots that the atoms start fluctuating and their fluctuations increase steadily up to almost 8ns before attaining an almost steady value of 1-1.1 nm around 20-30 ns and then it keeps on oscillating throughout the trajectory in the case of DNA backbones (black line). The time evolution of the RMSD of the bases exhibits a similar trend although the magnitude is slightly higher in this case (red line). The average values of the RMSD are calculated to be 0.8599 (±0.2752) nm and 0.8429 (±0.2015) nm for DNA bases and backbones respectively. However, it can be clearly observed that the average RMSD for bases and backbones are very close and it might underline the demerits of the RMSD calculations as a primary tool for probing atomic fluctuations associated with the DNA conformation at 0.01M concentration of $MgCl_2$. It is to be noted that the atomic fluctuations observed in the presence of only 6 $Mg^{2+}$ is substantial and at such low salt concentration time evolution as well as the average RMSD between two fragments of an ss-DNA from its native conformation can't be significantly different from each other. At higher concentration of $MgCl_2$ (0.1M), however, the differences between the time evolution, distribution and average RMSD between the DNA bases and the backbones are more apparent (**see supporting information, FIG.S2**) indicating higher magnitudes of atomic fluctuations.

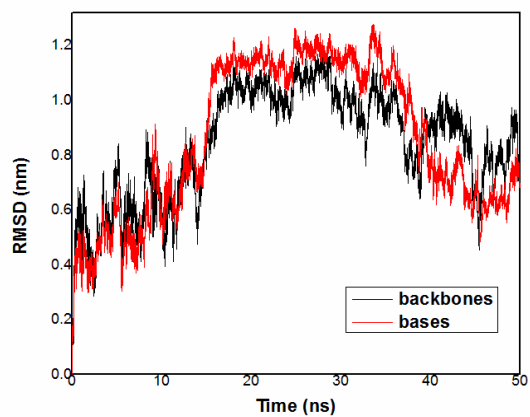

(a)

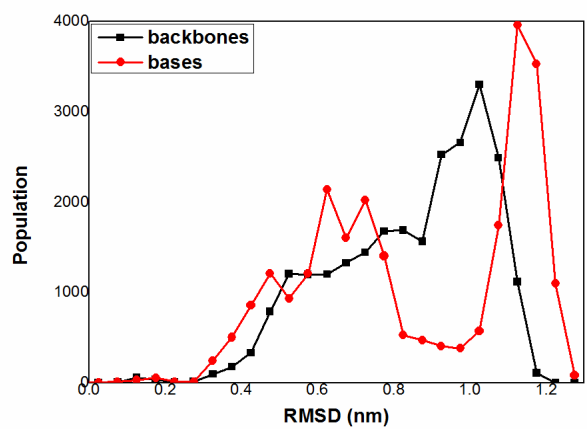

(b)

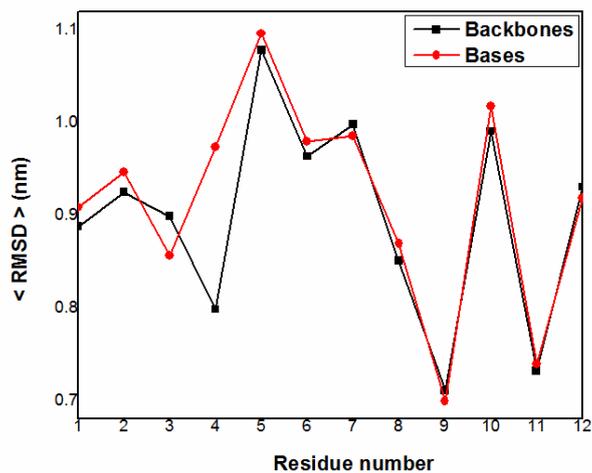

(c)

FIG.2. (a) Time evolution & (b) distribution of RMSD for DNA bases (red line) and backbones (black line). (c) Average RMSD values of bases and backbones for individual residues in the presence of 0.01M MgCl$_2$ from its native conformation.

Time evolution of the end-to-end distance which is one of the measures for the expansion or collapse of a chain has been depicted in **FIG.3.a.** The time evolution of $R_e$ clearly shows an onset of rapid decrease around 15 ns and it extends till 38 ns of the simulation time (**FIG.3.a**). The average $R_e$ values are 4.360±0.150 nm in the range of 0-15 ns and 1.063±0.021 nm in the range of 15-38ns of sampling time respectively. This is indicative of the collapse of the single stranded DNA. Beyond 38 ns and up to 50 ns of our simulation, $R_e$ ($<R_e>$ = 3.704±0.092 in this time frame) increases again which may be attributed to the incomplete structural collapse due to the lack of DNA counterion concentration and must not be confused with the overcharging induced DNA chain expansion since the integrated charge distribution $Q(r)$ of the system does not exceed 1 at 0.01M $MgCl_2$ concentration (**section III.D**). Similar insights can be drawn from the time evolution as well as the distribution of the radius of gyration of the simulated trajectory of the molecule in **FIG.3(c & d).** The average $R_g$ value (1.490±0.080 nm) up to 15 ns of simulation time gets reduced to 1.034±0.006 nm in the range of 15-38 ns is further indicative of the DNA chain compaction in this time frame. Similar to the time evolution of the DNA end-to-end distance, the radius of gyration also increases during the last 10 ns of sampling time ($<R_g>$ = 1.283±0.041 nm during 38-50 ns).

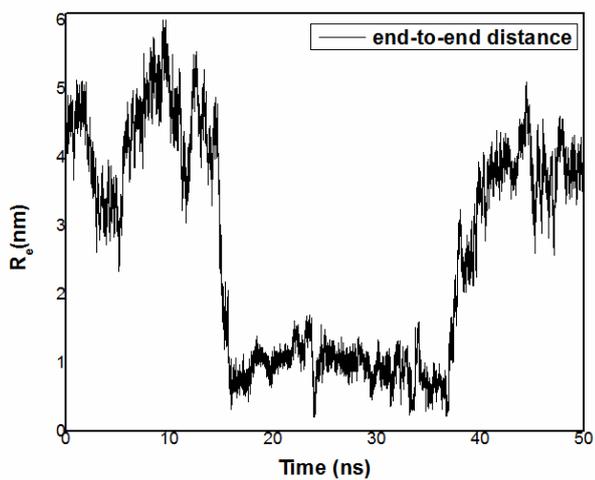

(a)

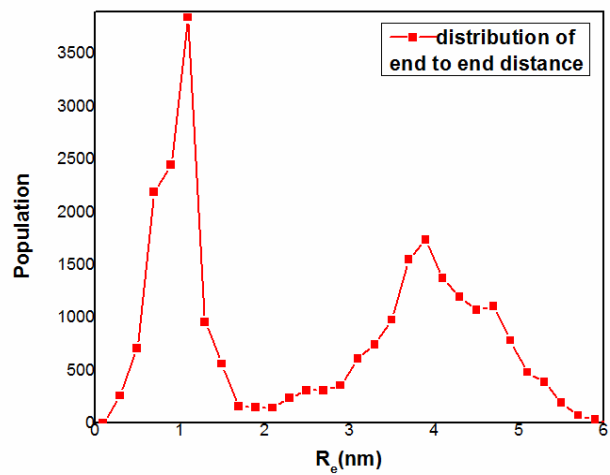

(b)

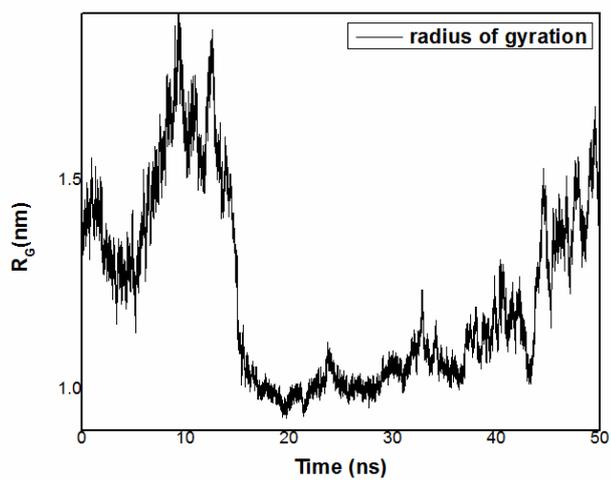

(c)

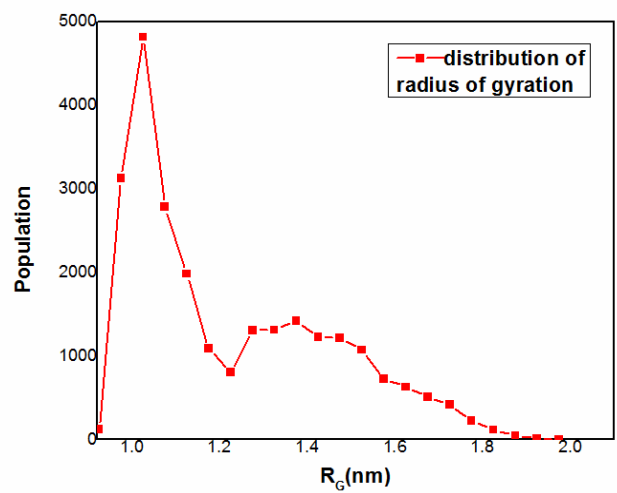

(d)

FIG.3. (a) Time evolution and (b) distribution of end-to-end distance. (c) Time evolution and (d) distribution of radius of gyration of the DNA molecule in the presence of 0.01M $MgCl_2$ concentration.

Time evolution of $R_e$ and $R_G$ establishes the collapse of the ss-DNA starting in around 15 ns. It is expected that the stacking between nucleobases would contribute to the collapse. An inherent feature of the native state of the DNA is the sequential stacking between consecutive base pairs which gets transformed into a completely heterogeneous distribution of nucleobases in the collapsed coil like state of the DNA where conformational rearrangements of the residues may lead to new combinations of base stacking that do not necessarily obey the original DNA sequence connectivity. This non sequential base stacking is supposed to be one of the major contributors in the collapsed DNA state[38]. We observed the emergence of new non sequential stacking motifs which disrupts the sequentially stacked ones to a considerable extent. To quantify this, we monitored the variation of the centre of mass distance ($R_S$) between any two residues, considering only ring atoms. We define two base residues to be stacked (sequentially or non-sequentially) if the calculated $R_S$ is ≤ 0.5 nm[39]. It is to be noted that by the term 'stacking' we imply the close proximity of any two DNA bases which resembles the π-π stacking between two aromatic rings in real world despite the fact that classical force fields cannot give a proper estimation of the π-π stacking since they cannot treat the π electrons or their polarizabilities explicitly. Therefore it is only an approximation to calculate the non-boned interaction energy parameters in connection with describing stacked DNA bases. While classical force fields fail to estimate the π-π stacking between DNA base pairs accurately even if the force fields are re-parameterized[40], there are quite a few works in which the non-bonded energy estimations are reported in connection with the stacking of nucleic acids using all-atom CHARMM force field[41,42]. Furthermore, it has also been shown that the extent of nucleic acid base flipping varies significantly with force fields (CHARMM27 and 36) though it is currently difficult to evaluate the accuracy of the employed force fields based on QM data[43]. Therefore, although the empirical CHARMM force field does not contain a specific energy term for π-π interaction, it can be assumed that the non-bonded energy terms described in the force field can be used in the context of the stacking of two DNA base pairs. In this study we calculate all possible $R_S$ distances of all residue combinations and some of them are displayed here. From **FIG.4** it is quite apparent that non sequential C1/G12 base pair remains stacked for almost 50% of the total simulation time while stacking in the case of the base pair C1/G4 exists only during the last 10 ns of the simulation time. However, most of the base pairs do not stay stacked throughout the remaining 50 ns simulation and the discontinuity of non sequential stacking may be attributed to inadequate overall DNA structural collapse due to the lack of sufficient counterion concentration. These non-sequentially stacked motifs contribute appreciably to the overall stabilization (both electrostatic as well as van der Waals energies) of the system in collapsed form. The average $R_S$ value for the above two non-sequentially stacked base pairs is found in the range of 0.42 – 0.49 (±0.07-0.09) nm for the two pairs while the corresponding average interaction energy terms comes out in the range of -5 (±0.18) kJmole$^{-1}$ for the pair C1/G12 and -12 (±1.18) kJmole$^{-1}$ for the C1/G4 pair. In addition, FIG.4 depicts the disruption of a sequentially stacked base pair T7/T8 which may apparently be a penalty for the emergence of the new non-sequentially stacked motifs.

Since most of our calculated parameters ($R_G$, $R_e$, $R_S$) do not exhibit a persistent trend with time, in order to conceive a better understanding of the dynamics of ss-DNA in the presence of divalent cations, additional MD simulations are performed at higher concentration of $Mg^{2+}$.

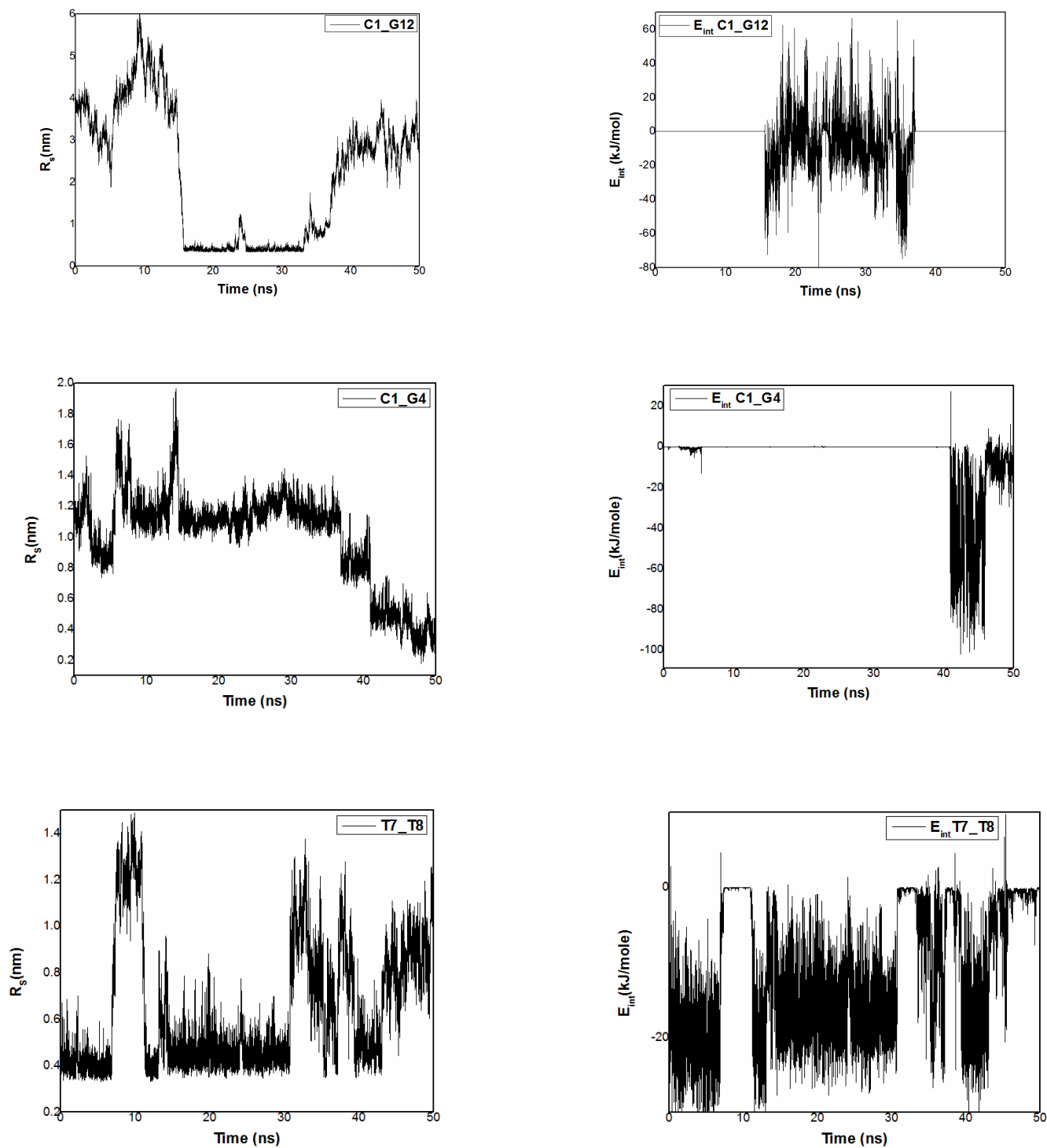

FIG.4. Variation of the centre of mass distance and interaction energy with time between two non-sequential and one sequential base pairs in the presence of 0.01M $MgCl_2$.

**B. Site specific ss-DNA interaction with water and ions**

**Structural arrangement of solvent molecules and ions**

In order to have a broader picture of the overall structural collapse of the DNA, it is equally important to probe the preference of positively charged $Mg^{2+}$ ions and water molecules to bind to the negatively charged phosphate backbone and the heterocyclic bases of the DNA molecule. The preferential binding of $Na^+$ ions in the AT sequence rich minor groove[44] and that of $Mg^{2+}$ in the deep major groove at the outer mouth of A-form nucleic acid duplexes[45] are quite well explored in literature. Similarly, we intend to probe the preferential occupancy of ions and solvent molecules in two binding pockets of the reference DNA, namely the bases and the backbones which in turn are responsible for the collapse of the ss-DNA. It is quite evident from **FIG.5.a** that oxygen atoms of water prefer to cluster around the oxygen atoms (black line) of the DNA bases (the $O_{base}$ group includes the O2, O4 and O6 atoms that belong to the DNA bases) while the radial distribution of water oxygen atoms plotted against its distance from the phosphate oxygen atoms (O1P and O2P in the $O_{backbone}$ group) reveals a much less intense peak (red line), implying the first hydration layer is more compact in the volume available at the interfaces of DNA bases, and not the backbones. If this is true, then the $Mg^{2+}$ density as calculated from the radial distribution should alternatively be higher for the backbones compared to the bases since the binding of divalent $Mg^{2+}$ with the negatively charged phosphate backbone is favourable. **FIG.5.b** indeed predicts the higher $Mg^{2+}$ affinity for the backbones compared to the bases. We calculate the average number of water molecules in the first coordination shell around the DNA oxygen atoms by integrating the g(r) for base and backbone oxygen atoms within 3.2 Å in **FIG.5.a**. On average three water molecule was found within the first coordination shell of DNA base oxygen atoms while only one water molecule is found around each DNA backbone oxygen atoms. This may be due to the presence of $Mg^{2+}$ ions which binds strongly to the phosphate backbone. This explains an increased water packing in the vicinity of DNA bases over that of DNA backbones. The above findings are consistent with both experimental[46] as well as simulation studies[47].

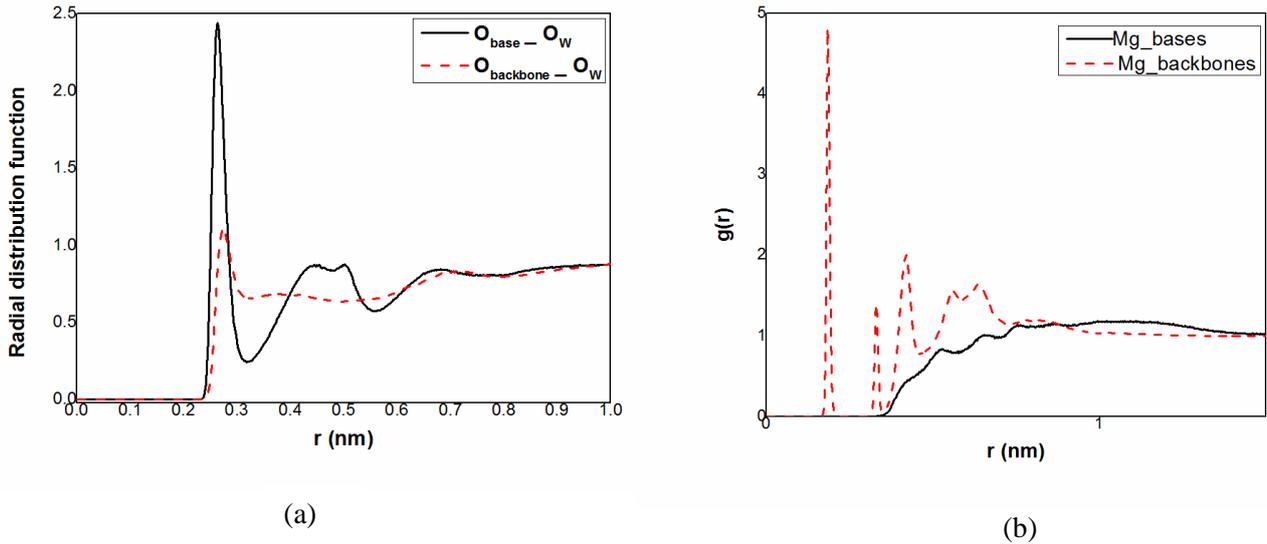

FIG.5. (a) Radial distribution function, g(r), of water oxygen atoms as a function of distance (in nm) from the phosphate oxygen atoms in the backbone (red line) and base oxygen atoms (black line) comprising the DNA. (b) g(r) of $Mg^{2+}$ ions as a function of its distances from the DNA bases (black line) and backbones (red line) in the presence of 0.01M $MgCl_2$.

## C. Effect of increasing ionic concentration on DNA conformation

An overview of the atomistic simulation of the DD dodecamer, followed by some experimental evidence[48-51] is useful in assessing the extent of ss-DNA structural collapse with the increase in the molar concentration of different DNA counterions. However, it will be interesting to observe the trend associated with the structural collapse of the DNA molecule at a series of higher concentrations of $Mg^{2+}$ since it is well known that polyelectrolytes such as DNA tend to undergo charge inversion[52] in presence of high molar concentration of DNA counterions. This phenomenon is quite well-known in polyelectrolyte physics and termed as 'overcharging'. DNA being a polyelectrolyte should also undergo chain expansion as a consequence of moderate to extensive overcharging[53]. Overcharging of polyelectrolytes in the presence of salts has been addressed by a number of research groups. Kundagrami and co-workers have shown that ion bridging by divalent counterions leads to a first-order collapse of simple polyelectrolytes in modest presence of divalent ions and the overcharging requires a minimum Coulomb strength[54]. Hsiao et al. used a combination of computer simulations and electrophoretic mobility studies to explore the charge inversion of a polyelectrolyte chain at high tetravalent salt concentration and have shown that the re-dissolution of polyelectrolyte condensates is not necessarily related to charge inversion[55]. Deserno et al. have shown the occurrence of overcharging for a model rod-like polyelectrolyte immersed in monovalent or divalent salt[56]. Furthermore, the effect of asymmetric salt and increasing macro ion size on charge inversion was demonstrated by Tanaka[57]. In order to demonstrate the impact of the overcharging in the case of ss-DNA, MD simulations are carried out in the presence of higher $MgCl_2$ salt concentrations, viz., 0.05, 0.1, 0.3 and 1M. The simulations are set up with the same initial configuration of the DNA (discussed in **section II**) using the same algorithms and the protocols discussed in **section II**.

In order to perceive a first-hand comparison of the extent of the structural collapse induced by $MgCl_2$ at different concentrations, viz., 0.05, 0.1, 0.3 and 1M, the time evolution of the end-to-end distance ($R_e$) and the radius of gyration ($R_G$) has been depicted in **FIG.6**. It immediately follows from the graph (**FIG.6.a**) that the value of $R_e$ starts decreasing sharply after 5 ns for 0.05M in comparison to 15 ns of the total sampling time for 0.1M $MgCl_2$. Once the end-to-end distance starts decreasing it hardly exhibits any jump for the rest of simulation time. This observation indicates that in the $MgCl_2$ concentration range of 0.05-0.1M, DNA structural collapse is monotonic. The average values listed in **TABLE 1** (see later) indicate a 20 % decrease in the $R_e$ and an almost 30 % decrease in the $R_G$ value on increasing the $Mg^{2+}$ concentration from 0.05M to 0.1M. However at higher concentrations of $Mg^{2+}$ the time evolution of $R_e$ and $R_G$ does not follow a monotonic pattern. The $R_e$ values in case of 0.3 and 1M $MgCl_2$ fluctuate and reach almost the initial value instead of an overall decrease with time (**FIG.6.a**). A similar non-monotonic variation of $R_G$ with time has been observed in the presence of 0.3 and 1M $MgCl_2$ (**FIG.6.b**). This anomaly can be attributed to the overcharging induced DNA chain expansion which is more dominant at a higher molar concentration of the DNA counterion. Overcharging influences the dynamics of polyelectrolytes in the presence of ions beyond a certain threshold concentration and it may be noted that the DNA chain swelling as a consequence of significant overcharging is consistent with the ion-strength dependent persistence length calculation of RNA by Pollack et al[58]. The finding of this work that the dynamics of the DNA is dominated by overcharging beyond 0.1M concentration of a divalent cation such as $MgCl_2$ is in good agreement with the work of both Wang et al[20] as well as that of Pai-Yi-Hsiao[18].

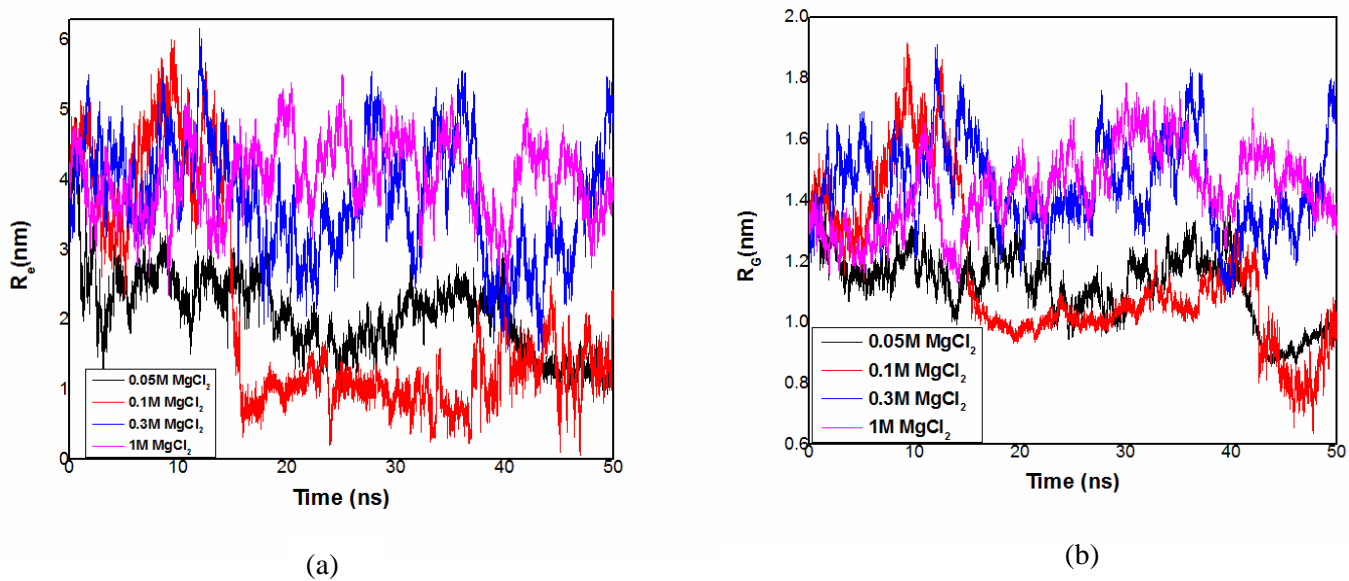

FIG.6. Time evolution of (a) end-to-end distance and (b) radius of gyration of the ss-DNA molecule at different concentrations of $MgCl_2$, viz., 0.05, 0.1, 0.3 and 1M.

The structural collapse of the ss-DNA causes certain rearrangements of different DNA conformational motifs, often leading to stable non-sequential base pair stacking which contributes appreciably to the overall stabilization of the DNA in its collapsed form. This kind of additional stacking interactions is thought to be dependent on ion concentration. We have already demonstrated the existence of such structural motifs at 0.01M $MgCl_2$ concentration. However, such a low concentration of $Mg^{2+}$ is not adequate for causing an overall collapse and the stacked base pairs (both sequential as well as non-sequential) deviate significantly till the end of the 50 ns simulation time (**FIG 4**). In order to investigate the stability of the stacked base pairs and the impact of overcharging induced DNA chain swelling at higher concentrations of $Mg^{2+}$ we have monitored the variation of $R_S$ (discussed earlier) between all possible base pair combinations with simulation time at different concentrations, viz., 0.05, 0.1, 0.3 and 1M of $MgCl_2$. We have scanned all possible combinations of non-sequentially stacked base pairs among which the one (G10-G12) that gets stacked in all of the above four concentrations and demonstrates the non-monotonic pattern of a divalent salt induced ss-DNA conformational changes have been depicted in **FIG 7** and **FIG 8**.

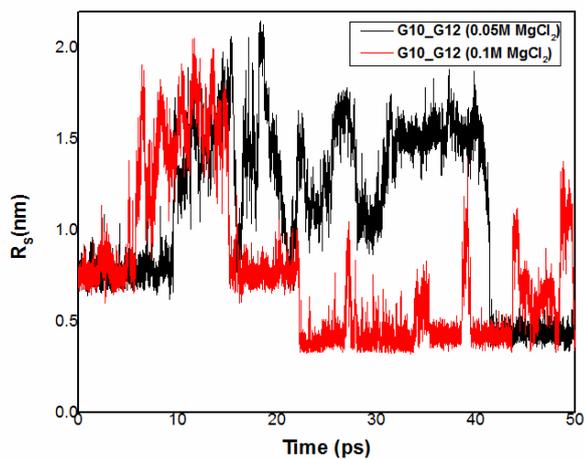
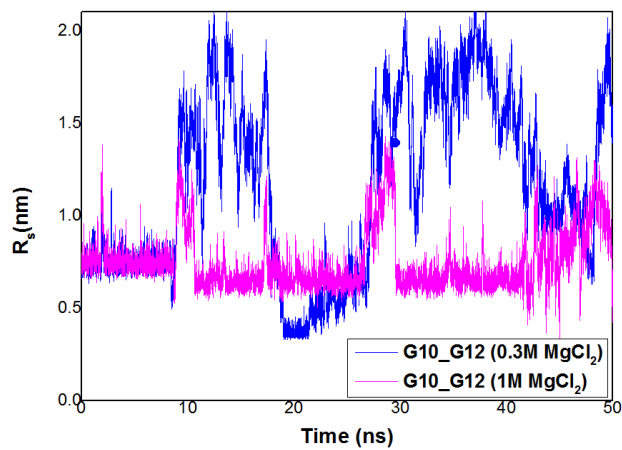

(a)                      (b)

FIG.7. Time evolution of the centre-of-mass distance ($R_S$) of a non-sequentially stacked base pair G10/G12 at (a) 0.05M & 0.1M and (b) 0.3M & 1M $MgCl_2$ concentration.

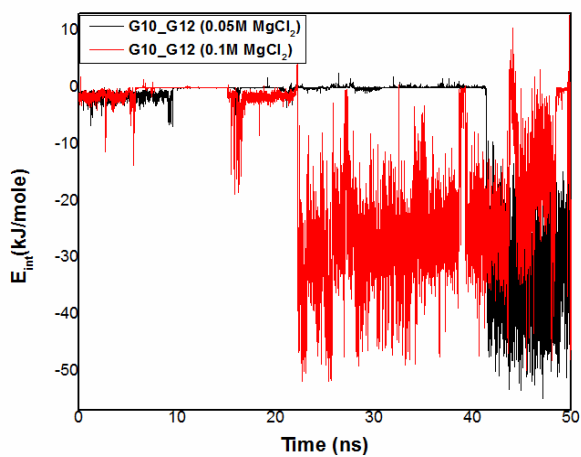
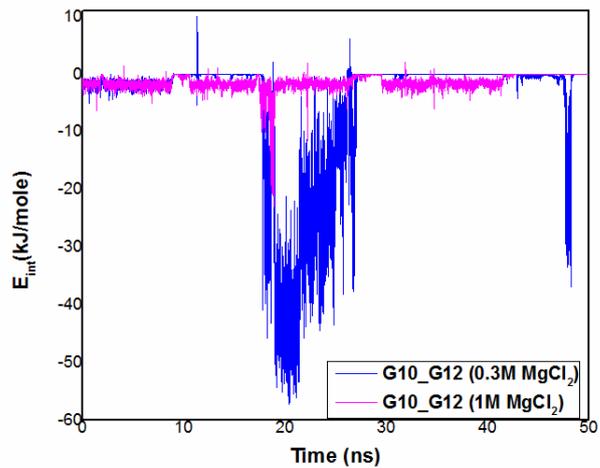

(a)                      (b)

FIG.8. Variation of the energy contribution of a non-sequentially stacked base pair G10/G12 at (a) 0.05 & 0.1M and (b) 0.3M & 1M $MgCl_2$ concentration with time.

It is evident from **FIG.7.a** that the lifetime of the non-sequentially stacked base pair G10 / G12 gets strengthened on increasing the $Mg^{2+}$ concentration from 0.05M to 0.1M. This pattern gets altered in **FIG.7.b** where a further increase in the molar concentration of the salt (0.3M) eventually destabilizes the same non-sequentially stacked base pair (G10/G12). However, the simulations do not provide an estimation of lifetime, which is commonly perceived as an ensemble average. Only one transition from a stacked to an unstacked DNA conformation is observed here per trajectory. A further increment in the concentration of the salt (1M) ultimately results in almost no stacking of the above base pair (**FIG.7.b**).The energy contributions (both electrostatic as well as non-bonded) of the non-sequentially stacked base pair G10/G12 are found in excellent agreement (**FIG.8.a & FIG.8.b**) with the time evolution of its $R_S$ (**FIG.7.a and FIG.7.b**) at different $Mg^{2+}$ concentrations, viz., 0.05, 0.1, 0.3 and 1M. The non-monotonic pattern of the stability of a non-sequentially stacked base pair as a function of $MgCl_2$ concentration is similar to the time evolution of $R_e$ and $R_G$ (**FIG.6.a** and **FIG.6.b**) of the ss-DNA and seems compatible with our assumption that overcharging predominates at higher molar concentrations of $Mg^{2+}$ (0.3 and 1M) and the discrepancies in the different measures of DNA chain folding ($R_e$, $R_G$ and $R_S$) are presumably due the effect of DNA chain expansion caused by overcharging.

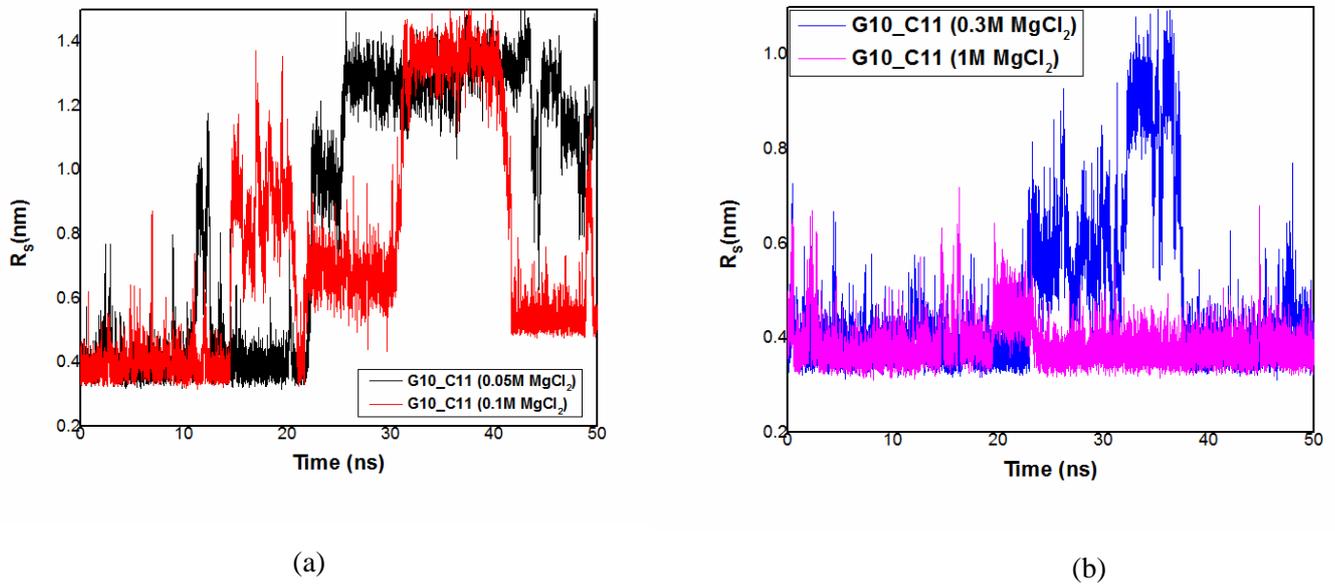

FIG.9. Time evolution of the centre-of-mass distance ($R_S$) of a sequentially stacked base pair G10/C11 at (a) 0.05M & 0.1M and (b) 0.3M & 1M $MgCl_2$ concentration.

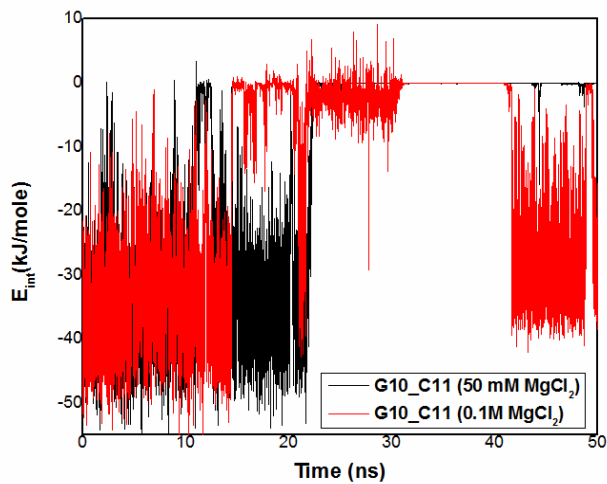 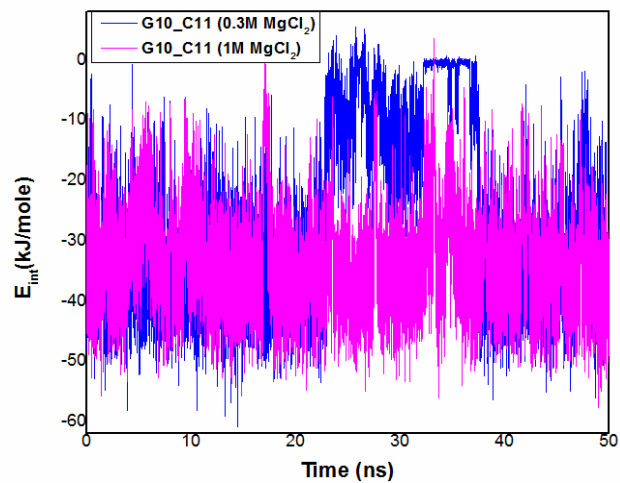

(a)

(b)

FIG.10. Variation of the energy contribution of a sequentially stacked base pair G10/C11 at (a) 0.05M & 0.1M and (b) 0.3M & 1M MgCl$_2$ concentration with time.

The emergence of a non-sequentially stacked motif is expected to influence the stacking of its sequential counterpart. In order to investigate this, we have monitored the time evolution of the R$_S$ of a sequentially stacked base pair G10/C11 at different concentrations. It is evident from **FIG 9.a** that the regular sequential arrangement of the base pair G10/C11 gets most disrupted at 0.05M and 0.1M concentration of MgCl$_2$, apparently a penalty for the stacking between G10 and G12 which are not sequentially connected. Since the stability of the non-sequentially stacked G10/G12 is found to be quite low (**FIG 7.b**) at 0.3 and 1M concentration of MgCl$_2$, the sequential stacking between G10 and C11 is observed almost throughout the entire 50 ns simulation time (**FIG 9.b**). The corresponding energy contributions are consistent with the sequentially connected G10/C11 pair getting disrupted at lower Mg$^{2+}$ concentrations (0.05M and 0.1M) and realigned at higher molar concentrations (0.3 and 1M) of MgCl$_2$ (**FIG 10.a** and **FIG 10.b**). A correlation between the average distance/energy parameters (**TABLE 1**) and the net charge distributions (**section III.D**) of the ss-DNA at different molar concentrations of $Mg^{2+}$ is helpful in understanding the importance of overcharging in the dynamics of an ss-DNA molecule in the presence of a divalent cation.

| Concentration (M) | $R_e$ (nm) | $R_G$ (nm) | $R_S$(G10/G12) (nm) | $E_{int}$ (G10/G12) (kJ mole$^{-1}$) |
|---|---|---|---|---|
| 0.05 | 2.508 (±0.118) | 1.167 (±0.1922) | 1.117 (±0.441) | -6.250 (±3.200) |
| 0.1 | 2.093 (±0.587) | 1.132 (±0.114) | 0.773 (±0.411) | -13.203 (±5.580) |
| 0.3 | 3.387 (±0.706) | 1.287 (±0.010) | 1.163 (±0.475) | -1.320 (±0.323) |
| 1.0 | 3.849 (±0.154) | 1.453 (±0.745) | 0.959 (±0.383) | -0.923 (±0.380) |

TABLE 1. The average distance/energy parameters calculated at different concentrations of $MgCl_2$ along with their standard deviations in parentheses.

## D. Ion binding and overcharging

The above described ss-DNA structural collapse is strongly related to the microscopic ion binding properties. We calculate the radial distribution of the net charge Q(r) which corresponds to the total net ion charges within a distance r from the ss-DNA chain. It is evident from **FIG 11.b** that the increase in charges in the vicinity of the DNA chain results in a stepwise increase in its ion binding and as a consequence the value of Q(r) increases continuously at smaller distances from the DNA before warding off to 1 at distances far apart from the DNA. The magnitude of Q(r) exceeding 1 at 0.3 and 1M $MgCl_2$ concentrations is indicative of these two systems being overcharged significantly presumably leading to DNA chain expansion at the above two concentrations. This is consistent with Muthukumar & co-workers's prediction for polyelectrolyte gel swelling to a large extent in divalent salts[59] and the crucial role of the ion excluded volume on the properties of a polyelectrolyte in the presence of a tetravalent salt[55]. The variations of the average distance parameters $R_e$ and $R_G$ (**FIG 11.a**) are in good agreement with the net charge distribution of the system (**FIG 11.b**).

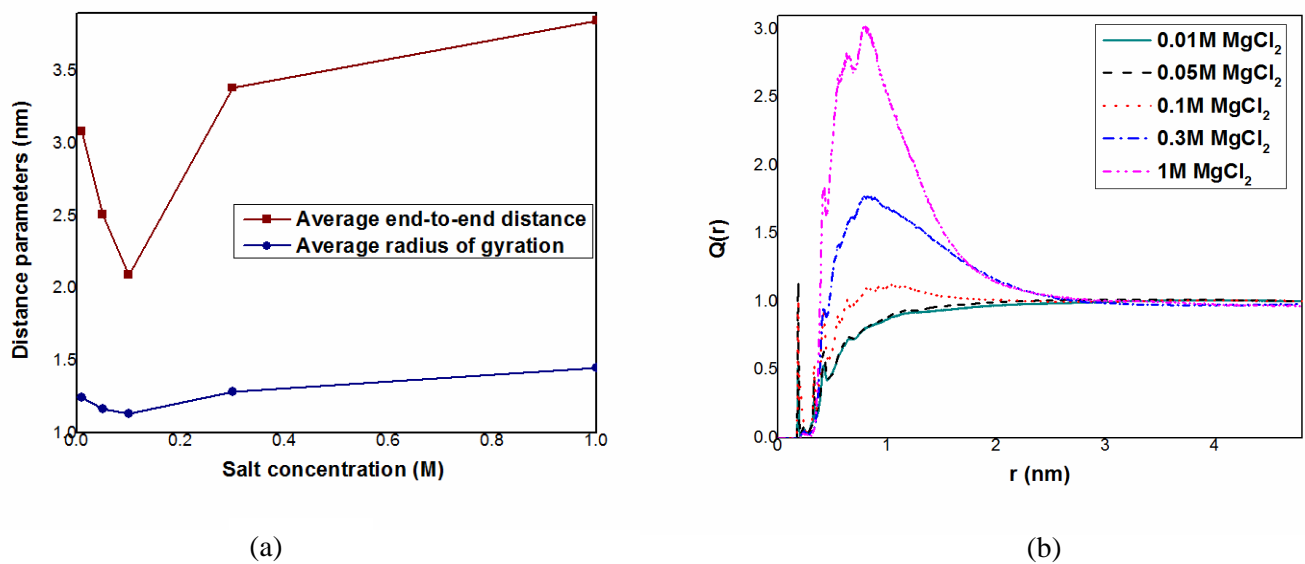

(a)          (b)

FIG.11 (a) dependence of distance parameters on ion concentrations. (b) Net charge distribution as a function of distance from the ss-DNA chain.

## IV. Conclusions

In this paper, using atomistic MD simulations we monitor the structural changes of an ss-DNA in the presence of a divalent salt at different concentrations. A preliminary evidence of the collapsed DNA structure is obtained by calculating the RMSD, end-to-end distance and radius of gyration. Our calculations indicate that the DNA bases are more fluctuating than the DNA backbones. The collapsed coil like conformation of the DNA molecule correlates with the formation of stable non-sequentially stacked base pairs C1/G12 and C1/G4 at 0.01M $MgCl_2$ concentration. However the short lifetimes of these stacked base pairs are due to inadequate DNA chain collapse. Atomistic MD simulations at higher molar concentrations of $MgCl_2$ revealed certain interesting facts. Interestingly, no definite trend in the time evolutions of distance /interaction energy parameters is observed due to significant overcharging induced DNA chain reswelling at 0.3 and 1M concentrations of $Mg^{2+}$. The findings of this work are contrary to that of Chakraborty and co-workers[16] where it has been shown that an ss-DNA molecule exhibits reduced end-to-end distance, followed by higher lifetime of the stacked base pairs on increasing NaCl concentration up to 500 mM. These qualitatively different results in the case of ss-DNA in the presence of the similar concentrations of $Na^+$ and $Mg^{2+}$ ions, agree reasonably well with the experimental findings by Pollack et al. that almost 20-40 times lower ionic strength of $MgCl_2$ than that of NaCl is sufficient to achieve the same chain compaction of ss-DNA and ss-RNA[58]. Caliskan et al[60] has also shown that the persistence length changes dramatically as RNA folds over a narrower concentration range of $Mg^{2+}$ as compared to $Na^+$.

Our simulations show that the extent of the DNA structural collapse increased monotonically from 0.05M to 0.1M under similar simulation conditions whereas the conformational deviations of the DNA are mostly dominated by overcharging induced DNA chain expansion at higher concentrations (0.3 and 1M) of $MgCl_2$. The magnitude of overcharging is quantified by calculating the net charge distribution as a function of distance from the ss-DNA (**section III.D**). The concept of the DNA getting overcharged beyond a certain threshold concentration is a precursor to the onset of DNA aggregation in the presence of monovalent and multivalent counterions[61]. According to the present work, the concentration of $MgCl_2$ at which the ss-DNA starts experiencing the effect of significant overcharging is 0.3M (**FIG. 11.b**) and this is consistent with the findings by Wang and co-workers[20].

We would also like to mention that the nature of the structural DNA collapse in the presence of different concentrations of a trivalent ion such as $Co^{3+}$ [62] along with a proper demonstration of the overcharging, would be appealing and worth investigating in future. Currently, we are working in the direction of quantifying ion specific overcharging and relate it with experimental observations using further atomistic simulations.

**Author information**

[*](R.C.) Telephone : +91-022-2576 7192  Fax : +91-022-2576 7192
Corresponding author
[*]Email: rajarshi@chem.iitb.ac.in

**Notes**:
The authors declare no competing financial interest.

## Acknowledgements

R. C thanks IIT Bombay – IRCC (Grant number: **12IRCCSG046**), DST-SERB **(SB/SI/PC-55/2013)** and S.G thanks CSIR, Govt. of India, for funding. R.C thanks Prof G. N. Patwari and S.G thanks Mr. M. K. Dixit for stimulating discussions.

# Ion assisted structural collapse of a single stranded DNA: a molecular dynamics approach

Soumadwip Ghosh[†], Himanshu Dixit[†], Rajarshi Chakrabarti[*]

*Department of Chemistry, Indian Institute of Technology, Powai, Mumbai – 40076, India.*

AUTHOR INFORMATION

**Corresponding author**

[*]Email: rajarshi@chem.iitb.ac.in

**Author Contribution**

[†]These authors contributed equally.

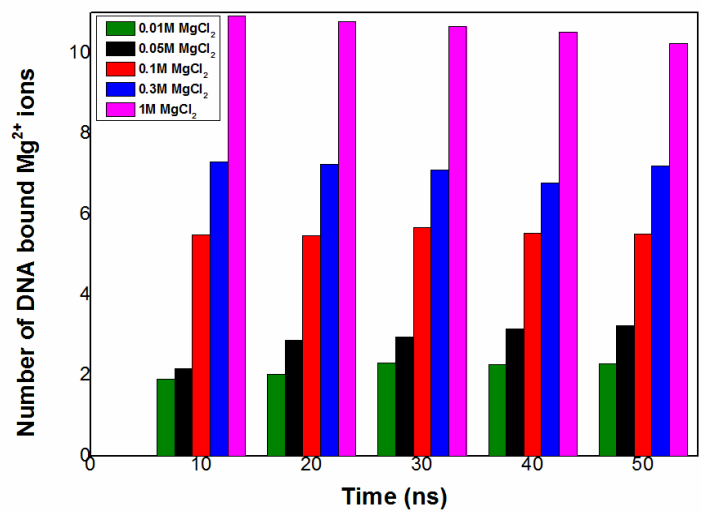

FIG. S1 Number of $Mg^{2+}$ ions binding the ss-DNA as a function of simulation time for each concentration of $MgCl_2$.

The time evolution of $Mg^{2+}$ ion distributions at different salt concentrations are determined by computing the radial distribution functions of $Mg^{2+}$ with respect to the centre of mass of the closest DNA atom. The first peak of such an RDF is observed around 5 Å and hence the number of 'bound' $Mg^{2+}$ is calculated within this cut-off. It is apparent from the above graph that the number of DNA bound cations increases steadily with increase in salt concentration which is consistent with the proposition of excess ion accumulation around an ss-DNA molecule in the present study. However, these numbers do not attain reasonable convergence at any salt concentration due to the inadequacy of statistical sampling time and the randomly introduced $Mg^{2+}$ ions into the system at the beginning of each simulation[30].

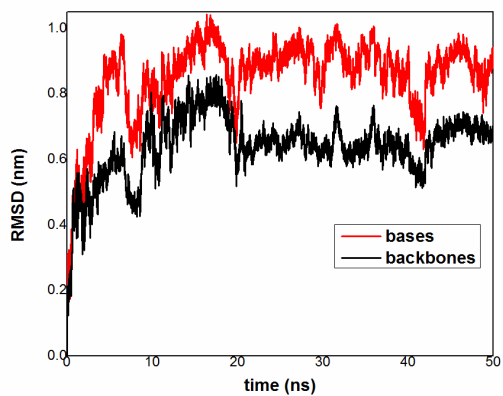

(a)

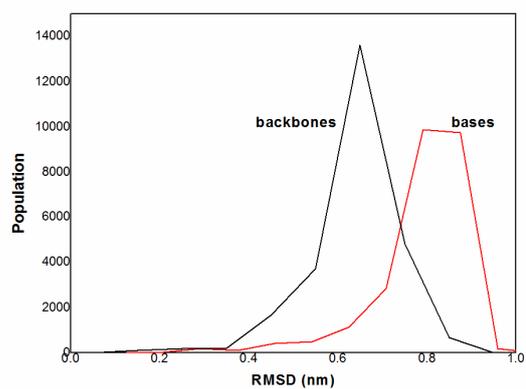

(b)

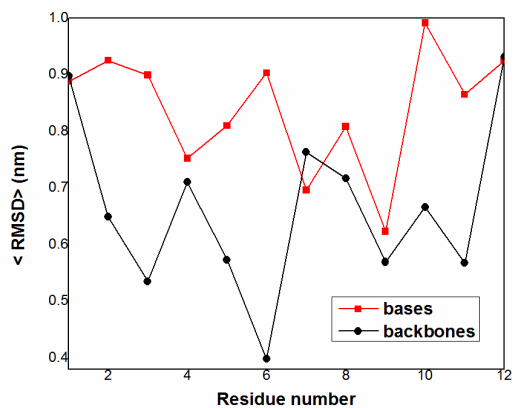

(c)

FIG. S2 (a) Time evolution & (b) distribution of RMSD for DNA bases (red line) and backbones (black line). (c) Average RMSD values of bases and backbones for individual residues in the presence of 0.1M MgCl$_2$.